\documentclass{aastex}
\usepackage{emulateapj5,apjfonts,psfig}

\slugcomment{Accepted for publication in {\it The Astrophysical Journal, August 2002}}
\begin{document}

\title{Is the EGRET source 3EG J1621+8203 the radio galaxy NGC~6251?}

\author{R. Mukherjee} 
\affil{Dept. of Physics \& Astronomy, Barnard College, Columbia University, New York, NY 10027}

\author {J. Halpern, N. Mirabal \& E. V. Gotthelf}
\affil{Dept. of Physics \& Astronomy, Columbia University, New York, NY 10027}

\begin{abstract}

We discuss the nature of the unidentified EGRET source 3EG~J1621+8203. In an 
effort to identify the gamma-ray source, we have examined X-ray images of the
field from {\sl ROSAT} PSPC, {\sl ROSAT} HRI, and {\sl ASCA} GIS. Of the
several faint X-ray point sources in the error circle of 3EG~J1621+8203, most
are stars or faint radio sources, unlikely to be counterparts to the EGRET
source. The most notable object in the gamma-ray error box is the bright FR I 
radio galaxy NGC 6251. If 3EG J1621+8203 corresponds to NGC 
6251, then it would be the second radio galaxy to be detected in high energy
gamma rays, after Cen A, which provided the first clear evidence of the 
detection above 100 MeV of an AGN with a large-inclination jet. 
If the detection of more radio galaxies by EGRET has been 
limited by its threshold sensitivity, there exists the exciting 
possibility that new high energy gamma-ray instruments, with much higher
sensitivity, will detect a larger number of radio galaxies in the future.   
\end{abstract}


\section{Introduction}

Surveys of the gamma-ray sky by the Energetic Gamma Ray Experiment Telescope 
(EGRET) on the Compton Gamma Ray Observatory (CGRO) in the 30 MeV to 10 GeV
energy range have 
revealed 271 point sources of gamma rays (Hartman et al. 1999). 
Of these the largest
group of identified sources are the active galaxies of the blazar class. 
Blazars are sources with high gamma-ray luminosity, which are 
characterized by flat radio spectra, rapid temporal variability at most
frequencies, and a high degree of optical 
polarization. Besides the blazars, only 
two other extragalactic sources have been detected by EGRET, namely the radio
galaxy Cen A, and the normal galaxy LMC. Most of the 65\% other sources
remain unidentified due to lack of convincing counterparts at other 
wavelengths. 

EGRET sources have large error contours, typically $\sim
0.5^\circ-1^\circ$, which makes identifications on the basis of position 
alone challenging. The 
principle method of identification of the EGRET sources relies on finding 
positional coincidences with flat-spectrum radio quasars (e.g. 
Thompson et al. 1995) or is based 
on the statistical evidence that blazars are the dominant population 
(e.g. Mattox et al. 1997; Mattox, Hartman \& Reimer 2001), relying in 
part on the sources' 5 GHz flux density. 
Often additional information from observations at other
wavebands are used to aid in the identification of the EGRET sources. Progress
has been made with recent studies in the X-ray band carried out for several 
EGRET fields (e.g. Roberts et al. 2001), and used to suggest counterparts for
a few unidentified EGRET sources, such as 3EG J2227+6122 (Halpern et
al. 2001a), 3EG J1835+5918 (Mirabal et al. 2000; 2001; Reimer et al. 2001), 
and 3EG J2016+2657 (Mukherjee et al. 2000; Halpern et al. 2001b). 

In this article we present gamma-ray and X-ray observations of one
unidentified EGRET source, 3EG J1621+8203. 
In particular, we examine the evidence in order to
see if 3EG J1621+8203 could be a second radio galaxy to be detected in high
energy gamma-rays by EGRET. 

\section{The Gamma Ray Observations}

3EG J1621+8203 is a high latitude source located at 
$l$=115.53, $b$=31.77, with a 95$\%$ confidence error radius of $0^\circ.85$
(Hartman et al. 1999). Recently, Mattox, Hartman, \& Reimer (2001) have 
generated elliptical
fits to the 95\% confidence contours for the Third EGRET Catalog (3EG) 
sources. This analysis yields an error ellipse with semi-major and semi-minor
axes of $0^\circ.97$ and $0^\circ.74$, respectively. Figure 1 shows the EGRET 
source position superimposed on the {\sl ROSAT} X-ray image that is described in 
\S 3.


3EG J1621+8203 is not very bright in high energy gamma rays,
and individual viewing periods yielded near-threshold detections by EGRET. In
the cumulative exposure from multiple EGRET viewings, 3EG J1621+8203 was 
clearly detected and the measured flux above 100 MeV was $1.1 \times 10^{-7}$ 
photon cm$^{-2}$ s$^{-1}$ (Hartman et al. 1999). 

The background-subtracted $\gamma$-ray spectrum of 3EG J1621+8203 was 
determined by dividing the EGRET energy band of 30 MeV
-- 10 GeV into 6 bins and estimating the number of source photons in
each interval, following the EGRET spectral analysis technique of
Nolan et al. (1993). The data were fitted to a simple power law model of the
form $F(E) = k(E/E_0)^{-\alpha}$ photon cm$^{-2}$ s$^{-1}$
MeV,$^{-1}$ where $F(E)$ is the flux measured at an energy $E$. The photon
spectral 
index $\alpha$ and the coefficient $k$ are the free parameters of the fit. The
energy normalization factor $E_0$ is chosen so that the statistical errors in
the power-law index and the overall normalization are uncorrelated. The fit to
the gamma-ray spectrum of 3EG 1621+6203 yielded a 
photon spectral index of 2.27$\pm$0.53 and is shown in Figure 2. The majority 
of the high energy photons for this source are in the 100-500 MeV band.

\section{The X-Ray Observations}

Archival X-ray imaging observations were available for the field of the EGRET
source 3EG J1621+8203. {\sl ROSAT} (Roentgen
Satellite) observations with the Position Sensitive Proportional Counter (PSPC)
in the range 0.2 -- 2.0 keV covered most of the error contour of 3EG
J1621+8203. Additional partial coverage was also available from the {\sl ROSAT} High
Resolution Imager (HRI) and the {\sl ASCA} (Advanced Satellite for  Cosmology
and Astrophysics) Gas Imaging Spectrometer (GIS). Historically, these fields
have been studied in X-rays because they contained the radio-loud active 
galaxy NGC 6251 (e.g. Sambruna, Eracleous, \& Mushotzky 1999; Turner et al. 
1997; Mack, Kerp \& Klein 1997; Birkinshaw \& Worrall 1993), and the gamma-ray
burst GRB 970815, which generated target-of-opportunity observations (Murakami
et al. 1997). 

\centerline{\psfig{file=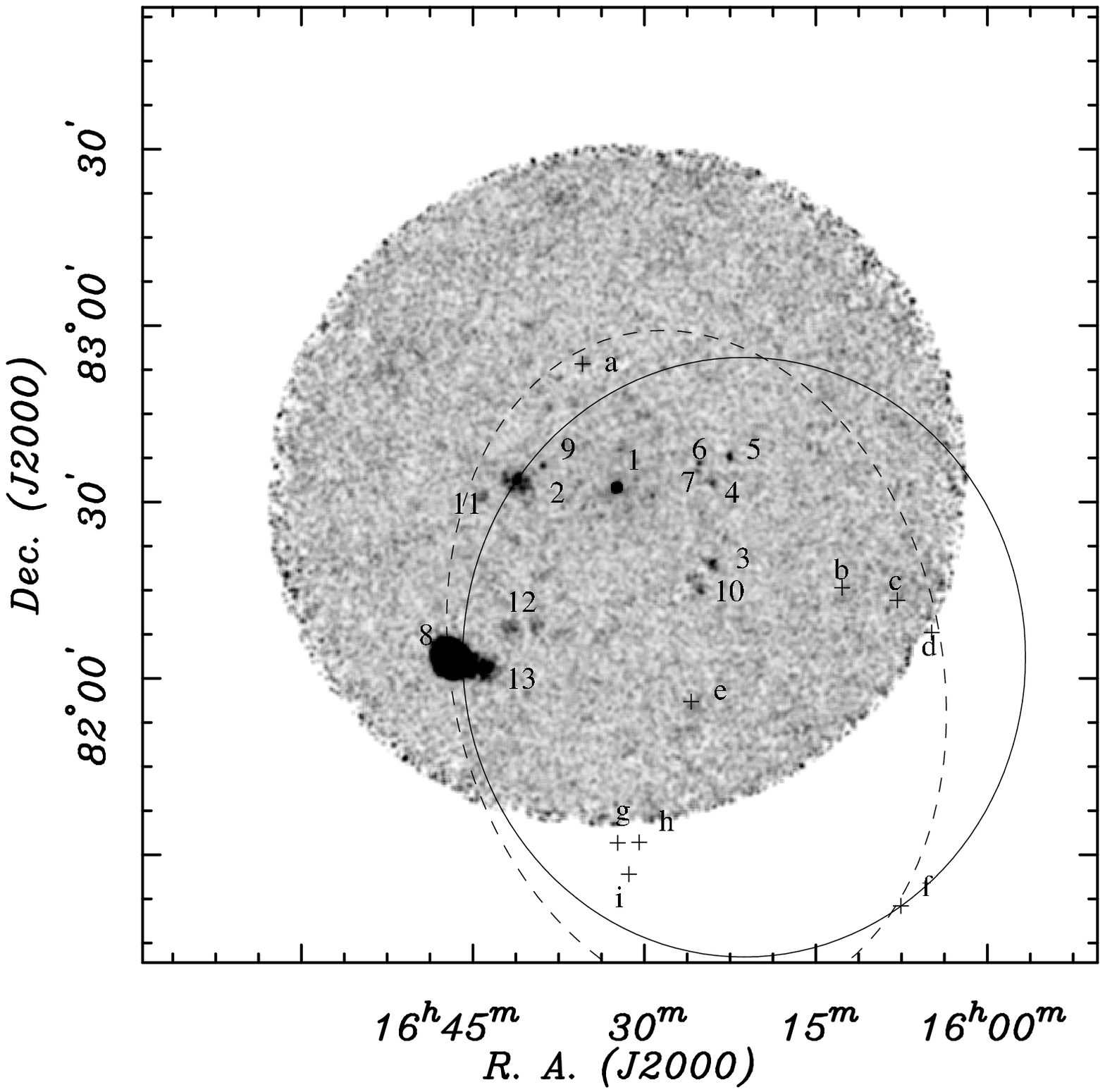,height=3.6in,bbllx=0pt,bblly=0pt,bburx=450pt,bbury=490pt,clip=.}}
\vspace{10pt}
{\footnotesize FIG. 1.--- {\sl ROSAT} PSPC soft X-ray image of 3EG J1621+8203,
exposure-corrected and smoothed with a $3\times3$ top hat kernel. The scaling
has been adjusted to highlight the faint emission. The 
circle corresponds to the $\sim 95$ \% confidence contours from the 3EG catalog
(Hartman et al. 1999). The dashed ellipse is the 95\% error contour derived 
from elliptical fits to the 3EG data (Mattox, Hartman, \& Reimer 2001). 
The numbers and
letters correspond to sources described in \S 3 and Tables 1 and 2. }
\bigskip

We have created {\sl ROSAT} PSPC image of the field of 3EG J1621+8203 
by co-adding exposure corrected sky maps of 14.7 ks of data taken during 1991
March 12-15, as shown in Fig. 1. The {\sl ROSAT} HRI and {\sl ASCA} GIS images
of the partial field of 3EG J1621+8203 are shown in Fig. 3. The figures also 
show the EGRET positions for the source, both as derived from the 3EG catalog,
as well as that from the elliptical fits to the EGRET data. 

In the archival {\sl ROSAT} and {\sl ASCA} data overlapping the EGRET 95\% 
error circle ellipse of 3EG J1621+8203 we find several faint X-ray point
sources. In addition, we have also searched the {\sl ROSAT} All Sky
Survey (RASS) catalog for sources in the field of 3EG J1621+8203, particularly 
in the regions not covered by the pointed {\sl ROSAT} PSPC and HRI data. The source
positions are marked in figures 1 and 3, and listed in Table 1. 
Count rates for the {\sl ASCA} and {\sl ROSAT} sources were obtained following
the method 
described in Gotthelf \& Kaspi (1998). Photons were extracted using a
$2^{\prime}$ radius aperture and the background contribution was estimated using a
large annulus away from the source. 

\centerline{\psfig{file=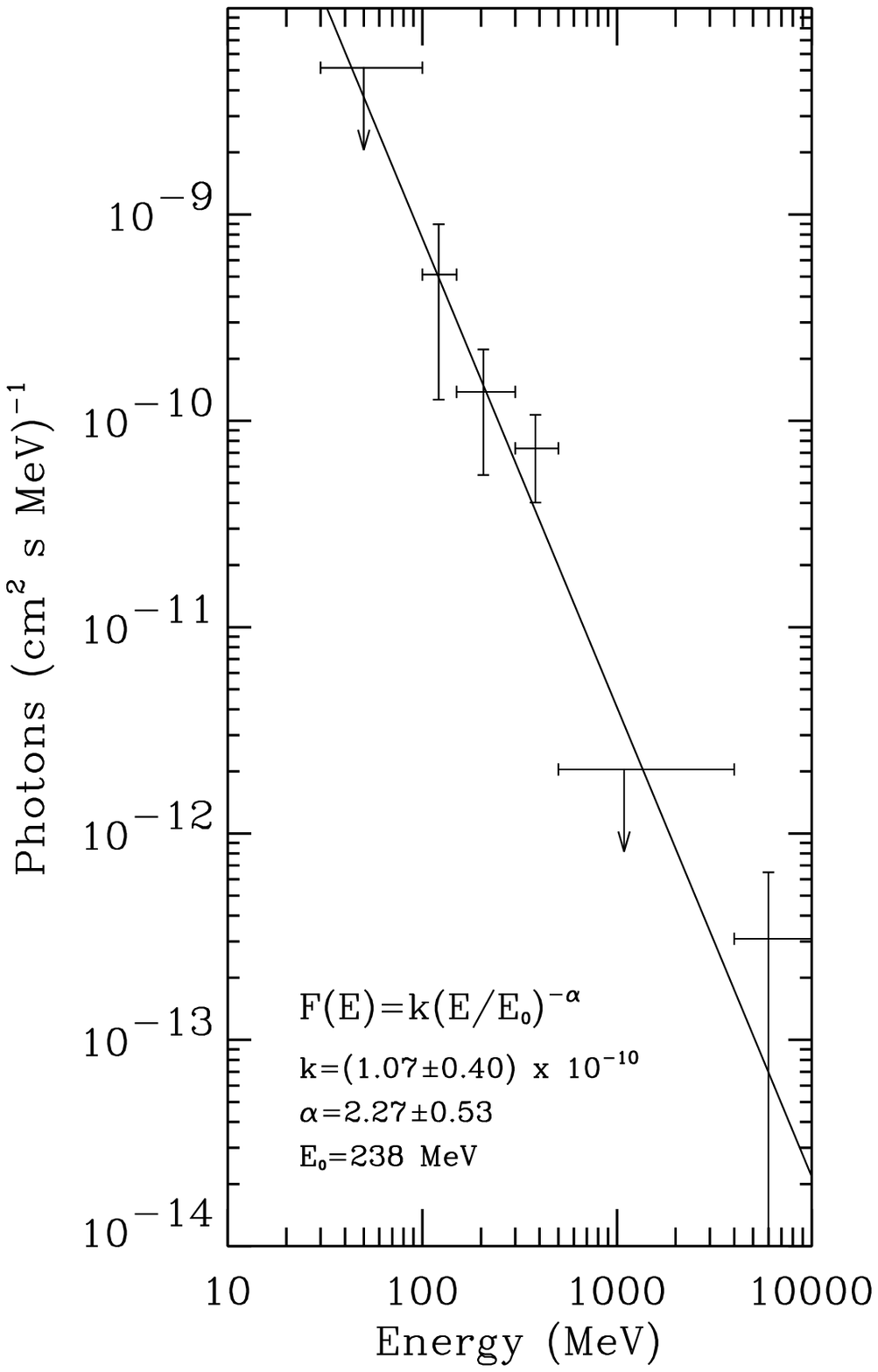,height=4in,bbllx=150pt,bblly=200pt,bburx=470pt,bbury=720pt,clip=.}}\vspace{10pt}
{\footnotesize FIG. 2.--- Photon spectrum of 3EG J1621+8203 in the the 30 MeV 
to 10 GeV energy range with superimposed fit to a simple power-law model. }
\bigskip

The {\sl ROSAT} sources are listed in Table 1 along with their background
subtracted count rates, detection significances, and hardness ratios. Sources
with count rates below $2.4\times10^{-3}$ cts/s are not listed in the table. 
The faintest source in Table 1 is a detection at the $3.0 \sigma$
level, with a background subtracted count rate of $0.0020\pm0.0006$ cts/s
in a 2.0$^\prime$ diameter circular aperture. For this analysis we report all
sources down to the $3\sigma$ detection level.             
The sources listed reached a minimum detectable intrinsic flux of $1.3\times
10^{-14}$ erg cm$^{-2}$ s$^{-1}$ in the 0.1 -- 2.4 keV band, 
assuming a power-law photon spectral index of 2.0, and no absorption (estimated
$N_H$ is $\sim 5\times 10^{20}$ cm,$^{-2}$ is expected to have a small effect). 

We have searched for optical and radio counterparts to the {\sl ROSAT} and {\sl
ASCA} point sources shown in figures 1 and 3. Most of the X-ray point sources
have possible optical counterparts, as listed in Table 1, and some are 
coincident with faint radio sources. Individual sources denoted by their source
numbers in Table 1 are described in the following section.  

\section{Notes on Individual Sources}

\begin{table*}
\begin{center}
\small
\caption{\centerline {X-ray sources in the fields of 3EG J1621+8203}}
\begin{tabular}{lcclccll}
\tableline
\#$^a$& RA & Dec& cts/s & HR$^c$& Optical/Radio & $\Delta^d$ & Suggested$^e$ \\
      &    &    &       &       &  Position     &            & Identification\\
\tableline
\tableline
1  & 16 32 32.5 & 82 32 15 & 0.102  & 0.92 & 16 32 32.0 +82 32 16 & 1'' & NGC 6251, Radio Galaxy \\
2  & 16 41 05.1 & 82 32 54 & 0.053  & 0.99 & 16 41 08.6 +82 33 09 & 17''  & 2E 1646.6+8238, \\
   &            &          &        &      &                      &       &  Galaxy Cluster, $z=0.26$\\
3  & 16 24 30.0 & 82 19 00 & 0.006  & 0.45 & 16 24 26.2 +82 18 53 & 10'' & USNO, $R=17.4$, $B=16.9$ \\
4  & 16 24 12.7 & 82 32 48 & 0.005  & 0.92 & 16 24 06.0 +82 32 49 & 13'' & USNO, $R=18.4$, $B=19.4$\\
5  & 16 22 32.6 & 82 37 06 & 0.008  & 0.66 & 16 22 34.0 +82 37 04 & 3.6''& $R=20.0$, $B= 20.2$\\
6  & 16 25 13.2 & 82 36 13 & 0.004  & 0.83 & 16 25 13.4 +82 36 21 & 8'' & USNO, $R=15.0$, $B=16.2$\\
7  & 16 25 27.3 & 82 34 49 & 0.002  & 0.91 & 16 25 28.2 +82 34 57 & 8'' & USNO, $R=18.9$, $B=19.1$\\
   &            &          &        &      & 16 25 27.7 +82 34 57 & 8''& NVSS, 46.2 mJy at 1.4 GHz\\
8  & 16 46 04.1 & 82 02 34 & 0.503  & 0.24 & 16 45 58.2 +82 02 14 & 25'' & HD 153751 (RS CVn type) \\
9  & 16 39 00.9 & 82 35 47 & 0.004  & 0.57 & 16 39 01.9 +82 35 49 & 3'' & USNO, $R=14.5$, $B=15.6$\\
10 & 16 25 29.6 & 82 14 39 & 0.005  & 0.74 & 16 25 31.4 +82 14 55 & 16'' & WN B1630.5+8221, \\
   &            &          &        &      &                      &      &  52 mJy at 0.325 GHz\\ 
11 & 16 44 12.6 & 82 30 07 & 0.009  & 0.33 & 16 44 07.8 +82 30 06 & 10'' & USNO, $R=18.3$, $B=18.0$\\
12 & 16 41 18.9 & 82 07 56 & 0.014  & 0.03 & 16 41 11.8 +82 07 53 & 15'' & USNO, $R=13.1$, $B=14.4$\\
13 & 16 42 58.1 & 82 01 02 & 0.016  & 0.13 & ... & ... & ...\\
14$^b$ & 16 03 24.5 & 81 42 19 & 0.016  & ... & 16 03 26.5 +81 42 21 & 4'' & BD+82 477, $B=10.05$, G0\\
15$^b$ & 16 06 52.0 & 81 30 28 & 0.003  & ... & .... & ... & $V > 24.3$, GRB 970815\\ 
16$^b$ & 16 13 17.3 & 81 23 33 & 0.010  & ... & 16 13 20.6 +81 23 33 & 8'' & USNO, $R=13.6$, $B=14.3$\\
17 & 16 37 20.6 & 82 07 36 & ...    & ... &  16 37 27.3 +82 07 07.5 & ... & Seyfert 1, $z=0.0402$ \\
\tableline
\end{tabular}
\end{center}

{\footnotesize $(a)$  Identifying number in {\sl ROSAT} PSPC image, Fig. 1.}
{\footnotesize $(b)$  {\sl ROSAT} HRI source, indicated in Fig. 3.} 
{\footnotesize $(c)$  Hardness ratio; $HR=(B-A)/(B+A)$, where $A$ is the count
rate in the energy range 0.1-0.4 keV, and $B$ is the count rate in the 
0.5-2.0 keV band (Voges, et al. 1999).} 
{\footnotesize $(d)$  Positional offset from the optical/radio position.}
{\footnotesize $(e)$  Most of the USNO objects are only suggested
IDs, based on positional coincidence and/or color.}\\
\end{table*}

1. {\sl  NGC 6251}: This is the bright FR I radio galaxy (Bicknell 1994; 
Urry \& Padovani 1995) at a
redshift of 0.0234 (implied distance 91 Mpc for $H_0=75$ km s$^{-1}$
Mpc$^{-1}$). NGC 6251 is the parent galaxy of an exceptional  
radio jet from pc to Mpc scale which makes an angle of $\sim
45^\circ$ to our line of sight (Sudou \& Taniguchi 2000). High dynamic range
observations of NGC 6251 with the VLBI show the presence of a bright core, and
an asymmetric jet, that implies relativistic beaming in this source 
(Jones et al. 1986). The VLBI observations are described further in \S 6. 
NGC 6251 has been studied in X-rays in the
past. A detailed investigation of the {\sl ROSAT} PSPC data was carried out by 
Birkinshaw \& Worrall (1993), who found the X-ray emission to be consistent 
with a point source at the position of the nucleus. Some extended X-ray 
emission surrounding the nucleus of NGC 6251 was noted, and a flat-spectrum
power-law component (photon spectral index $\Gamma\sim 1$) plus emission of a 
thermal plasma ($kT\sim 0.5$ keV) was found to fit the {\sl ROSAT} data best. 
The X-ray jet was studied in some detail by Mack, Kerp \& Klein (1997) who 
found that the X-ray emission in the jet is mostly due to bremsstahlung and 
line emission of hot, thin plasma, rather than synchrotron or inverse Compton
scattering. The {\sl ASCA} data for NGC 6251 were analyzed by Turner et
al. (1997) and Sambruna, Eracleous, \& Mushotzky (1999). The {\sl ASCA} 
spectrum was best modelled with a hard power law $\Gamma =1.83^{+0.21}_{-0.18}$
plus a thermal plasma component $kT=1.04^{+0.21}_{-0.18}$ keV 
that is harder than
observed in the {\sl ROSAT} data (Sambruna, Eracleous, \& Mushotzky 1999).
\medskip

\begin{table*}[t!]
\begin{center}
\small
\caption{\centerline{{\sl ROSAT} All Sky Survey Sources in the Field of 3EG J1621+8203$^a$}}
\begin{tabular}{cccccl}
\tableline
Number$^b$   &Source Name&  Cts/s & Optical Position & $\Delta^c$ & Magnitude$^d$ \\
\tableline
a &  1RXS J163525.2+825330 &  0.010 & 16 35 14.66 +82 53 31.7 & 20'' & $R=17.4$, $B=18.2$\\
b &  1RXS J161242.6+821525 &  0.009 & 16 12 46.30 +82 15 44.2 & 21'' & $R=18.7$, $B=19.7$\\
c &  1RXS J160752.0+821316 &  0.016 & 16 07 56.62 +82 13 03.2 & 16'' & $R=12.7$, $B=13.1$\\
d &  1RXS J160452.7+820748 &  0.022 & 16 04 56.66 +82 07 47.1 & 8''  & $R=15.7$, $B=17.4$\\
e &  1RXS J162554.6+815602 &  0.012 & 16 25 58.97 +81 55 56.7 & 11'' & $R=19.0$, $B=18.8$\\
f & 1RXS J160733.9+812118  &  0.014 & ...                     & ...    & ... \\
g & 1RXS J163027.4+813206  &  0.025 & 16 30 24.12 +81 32 09.6 & 8''  & $R=14.6$, $B=16.3$\\
h & 1RXS J163220.1+813201  &  0.015 & 16 32 23.19 +81 31 45.2 & 17'' & $R=12.4$, $B=13.5$\\
i & 1RXS J163121.3+812641  &  0.022 & 16 31 21.99 +81 26 43.7 & 3''  & $R=17.6$, $B=19.0$\\
\tableline
\end{tabular}
\end{center}
{\footnotesize $(a)$  {\sl ROSAT} all sky survey sources {\sl not} seen in the
pointed PSPC and HRI images.}
{\footnotesize $(b)$  Identifying number in the {\sl ROSAT} images (Fig. 1 \& Fig. 3).}
{\footnotesize $(c)$  Positional offset from the optical position.}
{\footnotesize $(d)$  Most of the USNO stars are only suggested
IDs, based on the fact that they are the nearest.} \\
\end{table*}

2. {\sl  RX J1641.2+8233}: A galaxy cluster, at a redshift of 0.26,
corresponding to the Einstein Imaging Proportional Counter (IPC) source 2E
1646.6+8238. This is an extended X-ray source with 90 galaxies within 
$3^\prime$ of the X-ray centroid, and all within 3 magnitudes of the three
brightest galaxies (Tucker, Tananbaum \& Remillard 1995). Another X-ray point 
source, 2RXP J164016.6+823203 is located close to the galaxy cluster. It is not
clear if this is a separate source or part of the cluster. 
\medskip

7. This is a possible QSO, based on its blue color and association with a weak
   radio source. 
\medskip

8. {\sl HD 153751}: Also known as SAO 2770, this is the
brightest X-ray source in the {\sl ROSAT} image of 3EG J1621+8203. It is a 
variable star of the RS CVn type (Eker 1992). It has a $B$ magnitude of 
5.098 and a $V$ magnitude of 4.22. Its spectral type is G5III. 
\medskip

10. {\sl WNB1630.5+8221}: A weak radio source with a flux density 
of about 58 mJy at 0.925 GHz. 
\medskip

14. {\sl BD+82 477}: Present only in a {\sl ROSAT} HRI image of the region, it
is also known as SAO 2659, and is a star of $B$ 
and $V$ magnitudes of 10.05 and 9.44, respectively, and spectral type G0
(Perryman et al. 1997). 
\medskip

17. {\sl 1AXG J163720+8207}: 
Using the MDM 2.4m telescope we have spectroscopically identified this source 
with a reddened Seyfert 1 galaxy at z=0.0402.  The coordinates
of this galaxy are (J2000) 16 37 27.3, +82 07 07.5.  (It should not
be confused with its physical companion, the galaxy IRAS F16419+8213,
which lies $1^{\prime}\!.3$ away and has the same redshift, but only 
H II region emission lines.)  This X-ray source is therefore unlikely
to be the EGRET source.

\medskip
Source 5 was imaged on the MDM 1.3 m, and an object with $R=20.0$ 
and $B=20.2$ was found in the field. Such color is typical of a QSO. 
The field of source 15 was empty 
in the optical down to the DSS limit. We have observed this field at the MDM,
and there is no optical counterpart to a $V$ magnitude limit of 24.3. Source 15
is the possible afterglow of GRB 970815 (Smith et al. 1999; Murakami et
al. 1997). The {\sl ASCA} and {\sl ROSAT} fields were observed as a result of the target of
opportunity generated by the gamma-ray burst. Source 15 is a weak X-ray source
seen both in the {\sl ASCA} and {\sl ROSAT} images, and is slightly outside the
error box of the GRB event, as reported by the RXTE/ASM (Smith et al. 1997). 
{\sl ASCA} observed the source three days after the GRB event, and the
flux of source 15 remained stable at $\sim  3 \times 10^{-13}$ erg cm$^{-2}$
s$^{-1}$ during the observations. The association of source 15 with GRB 970815 
remains uncertain. 

\centerline{\psfig{file=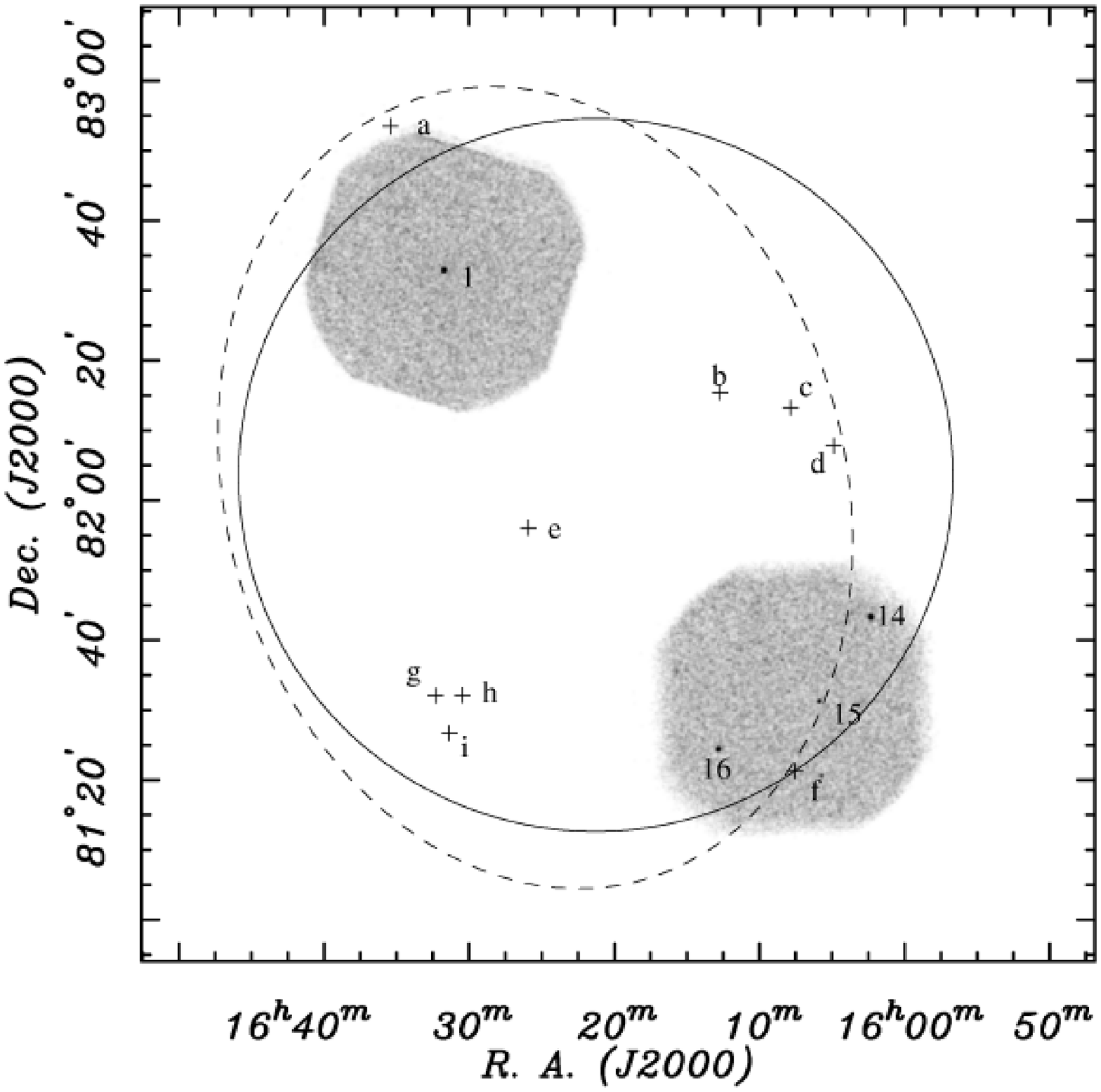,height=2.9in,bbllx=10pt,bblly=105pt,bburx=600pt,bbury=685pt,clip=.}}
\centerline{
\psfig{file=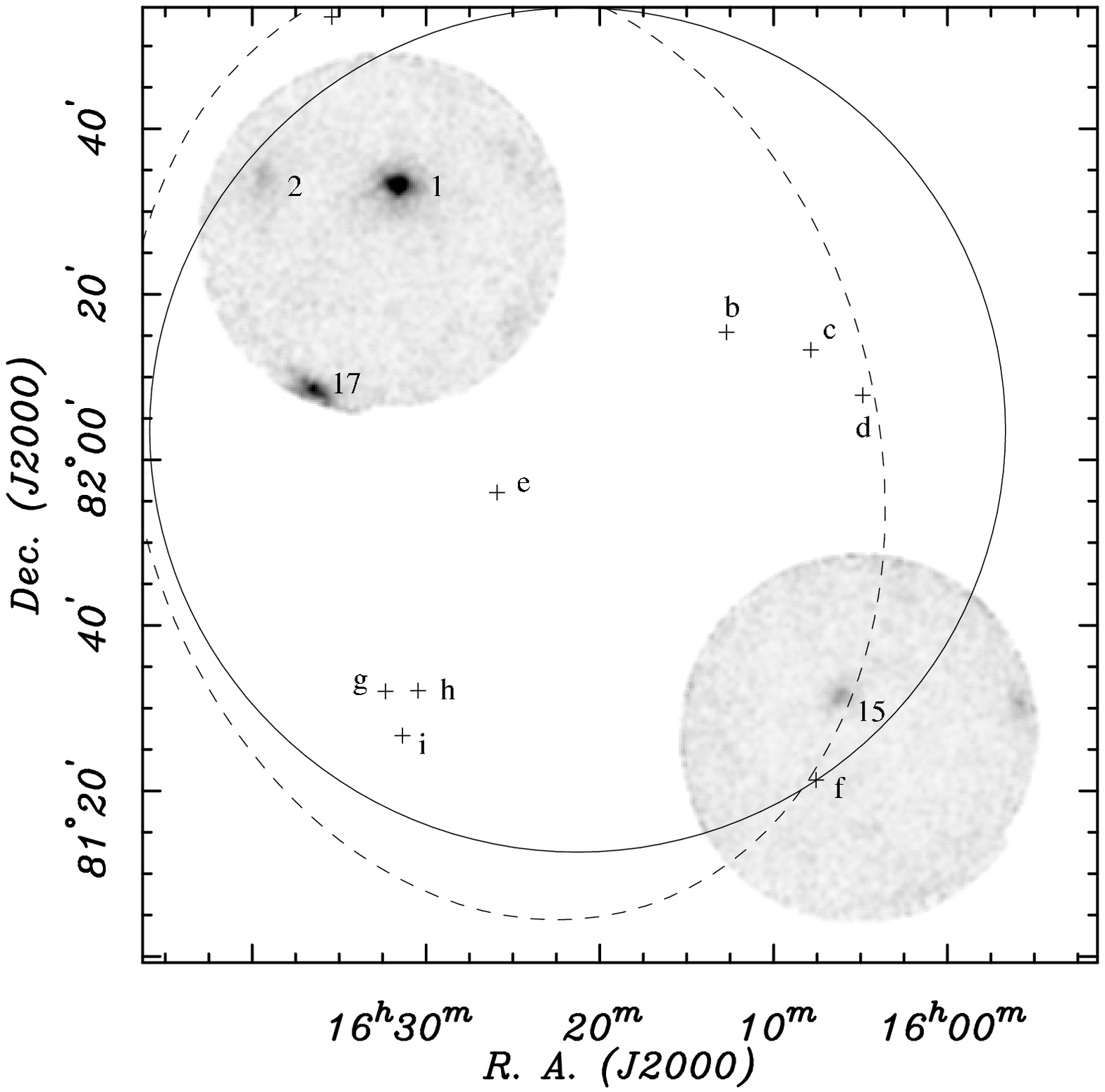,height=2.9in,bbllx=0pt,bblly=0pt,bburx=450pt,bbury=450pt,clip=.}}
{\footnotesize FIG. 3.--- (Top) {\sl ROSAT} HRI image and (Bottom) {\sl ASCA} image in the field
of 3EG J1621+8203. The circles corresponds to the $\sim 95$ \% confidence
contours from the 3EG catalog (Hartman et al. 1999). The dashed ellipses are the
95\% error contours derived from elliptical fits to the 3EG data (Mattox,
Hartman \& Reimer 2001). The numbers and  
letters correspond to sources described in \S 3 and Tables 1 and 2. }
\bigskip

The remaining numbered sources, except for
13, have positional
correspondences with objects in the USNO star catalog. Their $R$ and $B$
magnitudes are included in Table 1.  
\medskip

{\sl ROSAT All Sky Survey Sources (RASS)}: In addition to the point sources in
Table 1, Figs. 1 and 3 also show the locations of faint sources from the RASS
catalog (Voges et al. 2000) not detected in the pointed 
PSPC and HRI data. The RASS field covers the entire
EGRET error circle, even the areas not covered or detected by the pointed PSPC 
and HRI data sets. The RASS sources are marked with crosses and listed in Table
2. 
Except for the source marked ``f'' all the sources in
Table 2 are positionally coincident with objects from the USNO catalog. Source ``g''
was determined to be a dMe star from spectroscopic observation on the 
MDM 2.4m. 

\section{The Radio Observations}

We have searched the NRAO/VLA Sky Survey (NVSS) catalog (Condon et al. 1998) 
for possible 1.4 GHz radio counterparts to the X-ray point
sources. There were 111 radio sources in the field of 3EG J1621+8203 of 
which only 10 had integrated radio fluxes $> 100$ mJy. 
These are shown in Fig. 4 and listed in Table 3. Of 
these, $B_1$ is NGC 6251 and $B_2\cdots B_5$ probably correspond to emission
from the jet of NGC 6251. No significant X-ray emission is seen at the 
positions of the other sources and the NASA Extragalactic
Database (NED) reveals that these are steep spectrum radio sources. Mack, Kerp
\& Klein note another X-ray/radio coincidence east of source 7 at RA: 16 26 24.8
and Dec: 82 35 07. This source is 8C 1631+826, and has an X-ray flux density of 
$(9.8\pm2.9)\times 10^{-3}$ $\mu$Jy. 

It should be noted that that the most effective way to look 
for radio candidates for gamma-ray sources is to start not with the NVSS, but 
with a high-frequency radio catalog that is most likely to isolate the
flat-spectrum, blazar candidates.  The best such catalog in the 
northern hemisphere is the Becker, White, \& Edwards (1991) 4.85 GHz
survey.  However, it covers only declinations between 0$^\circ$ and 
+75,$^\circ$ which excludes this field. 

\begin{table*}[t!]
\tablenum{3}
\begin{center}
\caption{\centerline{Bright ($> 100$ mJy) NVSS Sources in the Field of 3EG J1621+8203}}
\begin{tabular}{ccccl}
\tableline
Number &  RA        &     Dec   & Flux (mJy)& Name \\
\tableline
B1 & 16 32 26.14 & +82 32 20.3&  802 & NGC 6251\\
B2 & 16 31 47.88 & +82 32 54.7&  144 & NGC 6251 jet \\
B3 & 16 30 51.13 & +82 33 45.6&  875 & '' \\
B4 & 16 29 25.21 & +82 35 34.1&  131 & ''\\
B5 & 16 27 42.31 & +82 42 26.1&  114 & ''\\
B6 & 16 28  9.52 & +81 50 21.3&  137 & 8C 1632+819\\
B7 & 16 26 43.74 & +81 58 50.8&  172 & 8C 1631+820\\
B8 & 16 24 11.83 & +82 09 14.1&  334 & 8C 1628+822\\
B9 & 16 21 24.93 & +81 35 34.1&  100 & WN B1621.7+8142\\
B10& 16 07 29.40 & +82 01 34.5&  314 & 8C 1612+821\\
\tableline
\end{tabular}
\end{center}
\end{table*}


\section{Discussion}

Our analysis of the archival X-ray data of the field containing 
3EG J1621+8203 reveals that the region contains several bright stars, weak 
radio sources, a radio galaxy, and a galaxy cluster. We note that unlike 
the majority of the identified EGRET sources, 3EG J1621+8203 lacks a 
radio-loud, spectrally  flat, blazar-like source catalogued within its error
circle that could be its potential counterpart. No radio counterpart search was
carried out by Mattox, Hartman and Reimer (2001) for this source, as it is in 
a region of sky not covered by the 5 GHz Green Bank survey. Of the notable 
X-ray sources in Table 1 the brightest is the RS CVn star, source \# 8. X-ray
emission from such stars is believed to arise from
coronal loop structures (Rosner et al. 1978), or perhaps as a result of
interaction between the mass transfer stream and the mass gaining star in a
binary system (Welty \& Ramsey 1995). RS CVn stars are not known to be
gamma-ray emitters, and thus we reject source \# 8 as a counterpart to the
EGRET source. Similarly, none of the other stars listed in Table 1 is likely to
correspond to the gamma-ray source. 

\bigskip
\centerline{\psfig{file=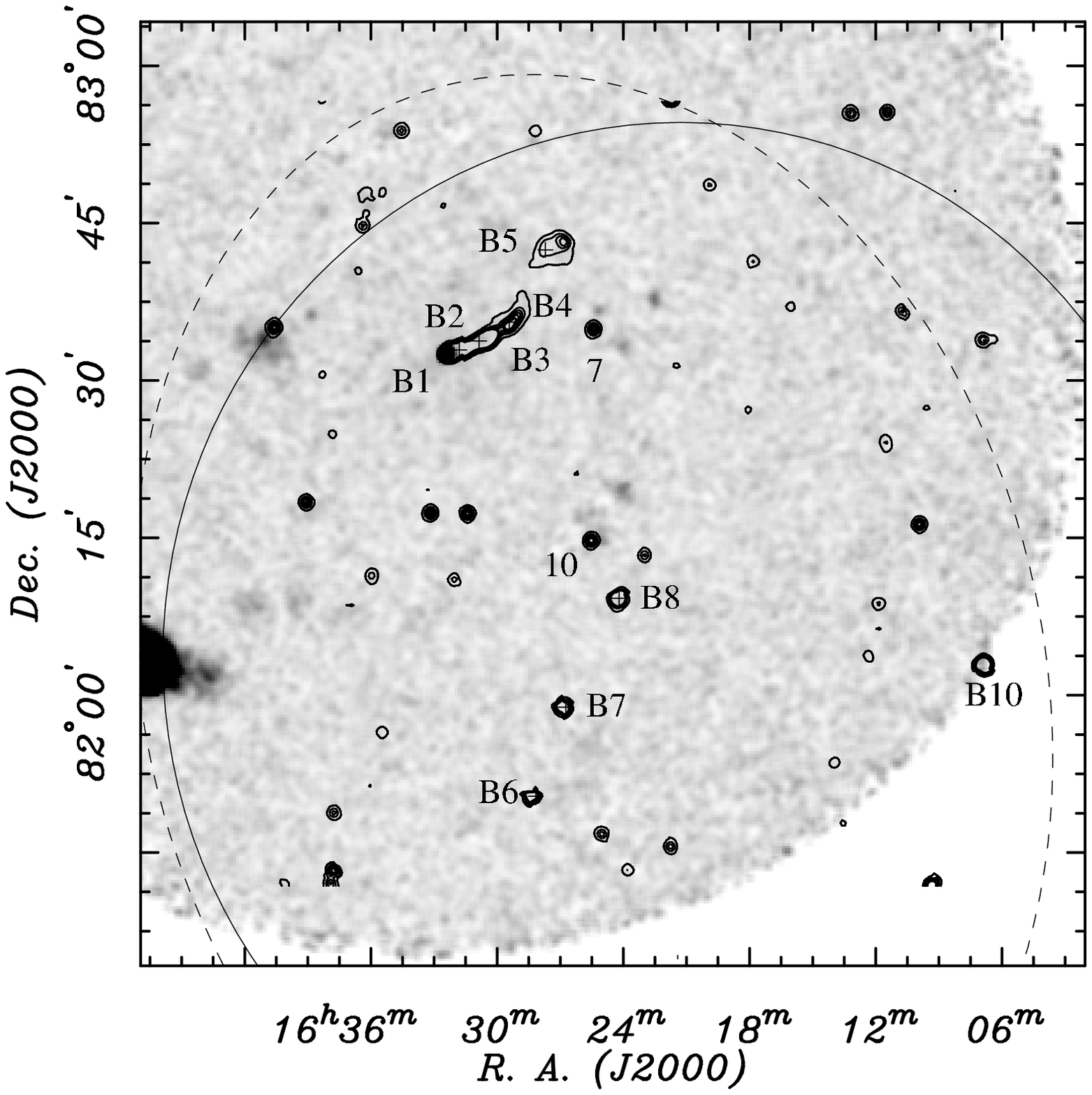,height=3.5in,bbllx=80pt,bblly=140pt,bburx=530pt,bbury=600pt,clip=.}}
{\footnotesize FIG. 4.--- Contour map of NVSS sources in the field of the EGRET
source 3EG J1621+8203 superimposed on the {\sl ROSAT} PSPC image (grey
scale). The source positions are listed in Table 3. Source B9, 
which is within the EGRET error contour but outside the X-ray image, is not
shown. X-ray sources 7 and 10 are coincident with weak radio sources listed in
Table~1. }
\bigskip

The two sources of interest in Table 1, however, 
are the galaxy cluster 2E 1646.6+8238 and the radio galaxy NGC 6251. 
The question of whether gamma-ray emission above 100 MeV is possible from X-ray
clusters is not yet resolved. It has been suggested that some of the
unidentified EGRET sources at high latitudes could be gamma-ray clusters,
perhaps a new population of gamma-ray emitting sources (Totani \& Kitayama
2000). Gamma-ray emission in clusters is presumably due to the inverse Compton 
scattering of the cosmic microwave background photons off high energy 
electrons accelerated in the shock waves produced during cluster
formation. However, Totani \& Kitayama (2000) note that there is no
statistically significant association of clusters from either the Abell catalog
(Abell, Corwin \& Olowin 1989) or the catalog of 
{\sl ROSAT} Brightest Cluster Sample (Ebeling et al. 1998) with the error
circles of the unidentified EGRET sources at high latitudes. In fact, the X-ray
surface brightness and surface number density of galaxies in gamma-ray clusters
are expected to be lower by a factor of about 200 and 30, respectively, than
those of ordinary clusters (Totani
\& Kitayama 2000), and the likelihood of finding gamma-ray clusters in past
X-ray and optical surveys is small. Brighter, closer sources of this class 
have not been previously detected by EGRET. In this case it is unlikely that 
the gamma-ray emission in 3EG J1621+8203 is due to the galaxy cluster 
2E 1646.6+8238, despite the positional coincidence. Perhaps deeper searches 
with more sensitive X-ray and gamma-ray instruments (e.g. GLAST) will resolve 
this issue conclusively in the future. 

Based on the present X-ray data, NGC 6251 appears to be the most likely
candidate identification for the EGRET source 3EG J1621+8203. In that case, NGC
6251 would be the second radio galaxy to be detected by EGRET. In the 3EG
catalog, Cen A (NGC 5128) is the only candidate detection of a radio galaxy at
energies above 100 MeV (Sreekumar et al. 1999). Its derived gamma-ray 
luminosity is weaker by a factor of $10^{-5}$ compared to the typical EGRET
blazar.  All other active galaxies identified with 
EGRET sources are blazars, which are believed to have 
nearly aligned jets along our line-of-sight. Cen A has a jet that is offset by
an angle of about $70^\circ$ (Bailey et al. 1986; Fujisawa et al. 2000). 
Cen~A happens to be the brightest and nearest 
radio galaxy ($z=0.0018$, $\sim 3.5$ Mpc). It has a 
double-lobed radio morphology, as evidenced from radio studies and shows the
presence of a one-sided X-ray jet, collimated in the direction of the giant
radio lobes. Cen A has a relatively weak radio luminosity $\sim 10^{40}$ ergs 
s$^{-1}$, and is classified as a FR I radio galaxy. In the past Cen A was
believed to be a misaligned blazar (Bailey et al. 1986). 

High dynamic range VLBI and VLA maps are available for NGC 6251 at 6, 13, 
and 18 cm, showing a bright core, a jet/counterjet brightness 
ratio of $\sim 80$ to 1 (Jones et al. 1986). The  fact that the source has a
nuclear structure and a large scale radio morphology, implies relativistic
beaming. Blazars detected by EGRET are compact radio sources,
with radio spectral indices $\geq -0.5$. The kpc-scale jet of NGC 6251 
has a slightly smaller spectral index (-0.64) (Saunders et al. 1981), but it 
could account for relativistic electrons and inverse Compton emission in this
source (Jones, et al. 1986). 

Figure 5 shows the spectral energy distribution 
of NGC 6251 assuming that it is  the counterpart to 3EG J1621+8203. The radio 
through optical data for the plot are from Ho (1999) and the references
therein. The {\sl ROSAT} data is from Worrall \& Birkinshaw (1994) who find 
that
90\% of the total PSPC flux comes from an unresolved component of diameter 
$\leq 4''$. The {\sl ASCA} data is from Sambruna, Eracleous \& Mushotzky 
(1999) and agrees with
the continuum model of Turner et al. (1997) who analyzed the data previously. 
We have also included VLBI and VLA data at 2.22, 6, and 18 cm (Jones et al. 1986). 
Similar to Cen A, the high energy gamma-ray emission from 3EG J1621+8203
represents a lower luminosity ($3\times 10^{43}$ ergs/s) than that of other
EGRET blazars (typically $10^{45}$ to $10^{48}$ ergs/s). 
Sreekumar et al. (1999) note that Cen A has a gamma-ray photon spectral index 
of $2.40\pm 0.28$, which is steeper than the average power-law spectrum from 
gamma-ray blazars ($2.15\pm 0.04$), and the spectrum of the extragalactic
gamma-ray background ($2.10\pm 0.03$). The gamma-ray spectral index of 
3EG J1621+8203 has a larger error (2.27$\pm$0.53), but it may be worth noting 
that it too is probably steeper than the average blazar spectrum. Unlike Cen A,
however, NGC 6251 has not been detected by either COMPTEL or OSSE. There is no
upper limit for this source in the first COMPTEL source catalog (Sch\"onfelder
et al. 2000).  

\bigskip
\centerline{\psfig{file=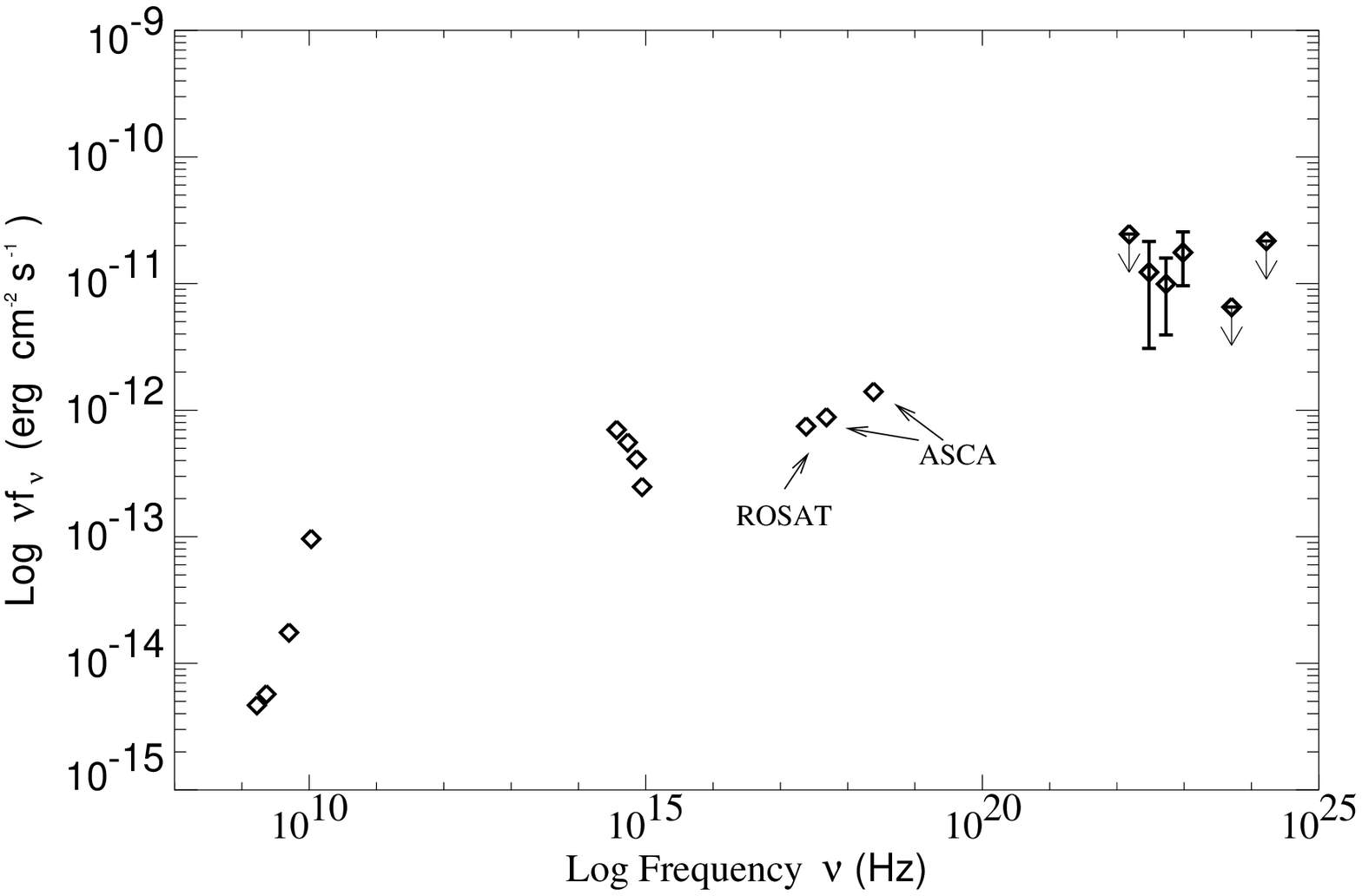,height=2.4in,bbllx=0pt,bblly=0pt,bburx=500pt,bbury=340pt,clip=.}}
{\footnotesize FIG. 5.--- Broad-band spectral energy distribution plot of NGC 6251,
assuming that it is the counterpart to 3EG 1621+8203. The radio through optical
data are obtained from a compilation by Ho (1999).}
\bigskip

The sensitivity of EGRET to off-axis emission from AGN, whose jets are 
pointed away from our line-of-sight, needs to be addressed. 
The total amount of scattered energy, $F_1$, as a function of viewing angle 
for an active galaxy may be estimated using the relation (see Dermer,
Schlickeiser, Mastichiadis 1992; Weferling \& Schlickeiser 1999), 
$$F_1(s,\mu^*_s)= D^{3+s}(1-\mu_s^*)^{(s+1)/2}\eqno (1)$$
where the gamma-ray flux seen by the observer is due to the scattered inverse
Compton emission of ambient low energy photons by highly relativistic 
particles in the jet. In this case, the particles are assumed to be electrons
and positrons, distributed in energy as a power-law with a spectral index of
$s$. $\mu_s^*$ is the cosine of the angle between the jet axis and the 
direction to the observer, and $D$ is the Doppler factor of the blob, defined 
as $D=\Gamma^{-1}(1-\beta\mu_s^*)^{-1}$, where $\beta c$ is the bulk velocity of the 
plasma. 

Figure 6 shows the decrease in scattered energy for off-axis 
emission, using the above relation, for different viewing angles, 
corresponding to two typical values of Lorentz factors ($\Gamma$) seen in 
blazars. The figure shows that a decrease in 
observer angle from $70^\circ$ (e.g. Cen A) to $45^\circ$ (e.g. NGC 6251) 
corresponds to an increase in the scattered energy by about a factor of 10, all
other things assumed equal. 
Cen A ($z=0.0018$) is the only source to be detected by EGRET with a large
inclination angle, presumably due to its proximity to Earth. NGC 6251 
($z=0.0234$) is much further away, but it is possible that the source is 
still detectable by EGRET due to its smaller jet angle. 

\centerline{\psfig{file=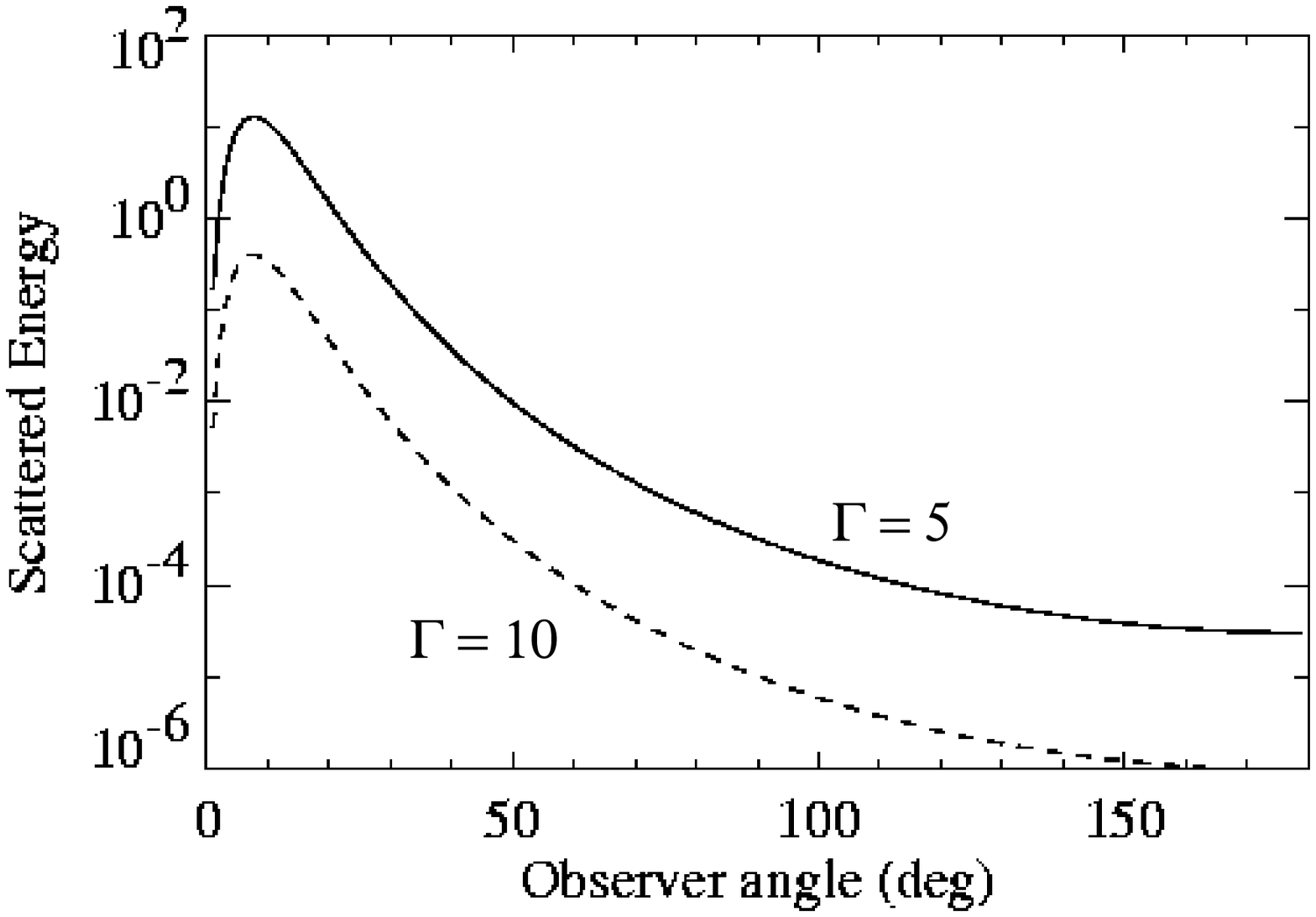,height=2.6in,bbllx=25pt,bblly=200pt,bburx=550pt,bbury=600pt,clip=.}}
{\footnotesize FIG. 6.---  Decrease in the observed emission from a blazar as a 
function of jet orientation with respect to the observer (Weferling \&
Schlickeiser 1999; Sreekumar 1999). }
\bigskip

The threshold sensitivity of EGRET ($> 100$ MeV) for a single 2-week 
observation was $\sim 3\times 10^{-7}$ photons cm$^{-2}$ s$^{-1}$ (Thompson et
al. 1993). Due to the intrinsically low luminosity of radio galaxies and the
limitations of EGRET's sensitivity, it is not surprising that many 
radio galaxies have not been detected as gamma-ray sources above 100 MeV thus 
far. It is very likely that more distant radio-loud AGN with 
intermediate inclination angles will be detected in the future with higher 
sensitivity gamma-ray instruments. 
FR I galaxies have been hypothesized to be the likely parent 
populations of BL Lac objects, which are believed to be beamed FR I 
galaxies (Padovani \& Urry 1990; Ghisellini et al. 1993). Since the number 
density of radio-loud FR I sources is nearly 
1000 times larger than FSRQs and BL Lac objects, the possibility that
such sources could form a new source class for future instruments like VERITAS 
(e.g. Weekes et al. 2000) or GLAST (e.g. Gehrels \& Michelson 1999) 
is an exciting one. NGC 6251=3EG J1621+8203 and Cen A=3EG~J1324-4314 could be
examples of such sources. 
In fact, there exists the likelihood 
that such ``misaligned blazars'' could contribute to the extragalactic 
gamma-ray background around 1 MeV (Stienle et al. 1998; 
Sreekumar et al. 1999; Watanabe \& Hartmann 2001). 
Gamma-ray sources not resolved by present-day detectors 
must contribute to the extragalactic gamma-ray background detected by 
EGRET in the 30 MeV to 100 GeV range. 
The contribution of misaligned blazars to the gamma-ray background has
been calculated by Weferling \& Schlickeiser (1999) where the authors modeled
the EGRET-detected extragalactic gamma-ray background. The misaligned blazars
clearly outnumber the aligned ones, but direct observation of these sources may
only be possible in the future with more sensitive gamma-ray instruments.  

\bigskip
\smallskip
This research has made
use of data obtained from HEASARC at Goddard Space Flight Center and 
the SIMBAD astronomical database. R.M. acknowledges support from NSF grant 
PHY-9983836. E.V.G. is supported by NASA LTSA grants NAG~5-22250. 


{\bf References}
\parindent 0pt

Abell, G. O., Corwin, H. G., Olowin, R. P. 1989, ApJS, 70, 1. 

Bailey, J., Sparks, W. B., Hough, J. H., Axon, D. J. 1986, Nature, 322, 150. 

Becker, R. H., White, R. L., Edwards, A. L. 1991, ApJS, 75, 1. 

Bicknell, G. V. 1994, ApJ, 422, 542. 

Birkinshaw, M., \& Worrall, D. M. 1993, ApJ, 412, 568. 

Condon, J. J., Cotton, W. D., Greisen, E. W., Yin, Q. F., Perley, R. A., 
Taylor, G. B., Broderick, J. J. 1998, AJ, 115, 1693. 

Dermer, C. D., Schlickeiser, R., Mastichiadis, A. 1992, A\&A, 256, L27. 

Ebeling, H., Edge, A. C., Bohringer, H., Allen, S. W., Crawford, C. S., 
Fabian, A. C., Voges, W., Huchra, J. P. 1998, MNRAS, 301, 881. 

Eker, Z 1992, ApJS, 79, 481. 

Fujisawa, K., Inoue, M., Kobayashi, H., Murata, Y., Wajima, K., et al. 2000, 
PASJ, 52, 1021. 

Gehrels, N., \& Michelson, P. 1999, Astropart. Phys., 11, 277. 

Ghisellini, G., Padovani, P., Celotti, A., Maraschi, L. 1993, ApJ, 407, 65. 

Gotthelf, E. V., Ueda, Y., Fujimoto, R., Kii, T., Yamaoka, K. 2000, ApJ, 543,
417. 

Gotthelf, E. V., \& Kaspi, V. M. 1998, ApJ, 497, L29.  

Halpern, J. P., Camilo, F., Gotthelf, E. V., Helfand, D. J., Kramer, M., 
Lyne, A. G., Leighly, K. M., Eracleous, M. 2001a, ApJ, 552, L125.  

Halpern, J. P., Eracleous, M., Mukherjee, R., Gotthelf, E. V. 2001b, ApJ, 551,
1016. 

Hartman, R. C., Bertsch, D. L., Bloom, S. D., Chen, A. W., Deines-Jones, P., 
Esposito, J. A., Fichtel, C. E., Friedlander, D. P., et al. 1999, ApJS, 123, 
79.  

Ho, L. 1999, ApJ, 516, 672. 

Jones, D. L., Unwin, S. C., Readhead, A. C. S., Sargent, W. L. W., Seielstad,
G. A., et al. 1986, ApJ, 305, 684. 

Mattox, J. R., Schachter, J., Molnar, L., Hartman, R. C., Patnaik 1997, 
ApJ, 481, 95. 

Mattox, J. R., Hartman, R. C., \& Reimer, O. 2001, ApJS, 135, 155. 

Mack, K. -H., Kerp, J., \& Klein, U. 1997, A\&A, 324, 870. 

Mirabal, N., \& Halpern, J. P. 2001, ApJ, 547, 137L. 

Mirabal, N., Halpern, J. P., Eracleous, M., Becker, R. H. 2000, 541, 180. 

Mukherjee, R., Gotthelf, E. V., Halpern, J., Tavani, M. 2000, ApJ, 542, 740. 

Murakami, T., Ueda, Y., Yoshida, A., Kawai, N., Marshall, F. E., et al. 1997, 
IAU Circular, 6732, 1. 

Nolan, P. L., Arzoumanian, Z., Bertsch, D. L., Chiang, J., Fichtel, C. E., 
Fierro, J. M., Hartman, R. C., Hunter, S. D., et al. 1993, ApJ,
409, 697. 

Padovani, P., \& Urry, C. M. 1990, ApJ, 356, 75. 

Perryman, M. A. C., Lindegren, L., Kovalevsky, J., Hoeg, E., Bastian, U., 
Bernacca, P. L., Crézé, M., Donati, F., et al. 1997, A\&A, 323L, 49. 

Reimer, O., Brazier, K. T. S., Carramiñana, A., Kanbach, G., 
Nolan, P. L., Thompson, D. J. 2001, MNRAS, 324, 772.  

Roberts M. S. E., Romani, R. W., Kawai, N. 2001, ApJS, 133, 451. 

Rosner, R., Tucker, W. H., Vaiana, G. S. 1978, ApJ, 220, 643. 

Sambruna, R. M., Eracleous, M., \& Mushotzky, R. F. 1999, ApJ, 526, 60. 

Saunders, R., Baldwin, J. E., Pooley, G. G., Warner, P. J. 1981, MNRAS, 197,
287. 

Sch\"onfelder, V., Bennett, K., Blom, J. J., Bloemen, H., Collmar, W., 
Connors, A., Diehl, R., Hermsen, W., et al. 2000, A\&AS, 143, 145. 

Smith, D. A., Levine, A. M. , Morgan, E. H., Wood, A. 1997, IAU Circular, 6718,
1. 

Smith, D. A., Levine, A. M., Bradt, H. V., Remillard, R., Jernigan, J., et al. 
1999, ApJ, 526, 693. 

Sreekumar, P., Bertsch, D. L., Hartman, R. C., Nolan, P. L., Thompson, D. J. 
1999, Aph, 11, 221. 

Stienle, H., Bennett, K., Bloemen, H., Collmar, W., Diehl, R., Hermsen, W., 
Lichti, G. G., Morris, D.,  et al. 1998, A\&A, 330, 97.  

Sudou, H., Taniguchi, Y. 2000, AJ, 120, 697. 

Thompson, D. J., Bertsch, D. L., Fichtel, C. E., Hartman, R. C., 
Hofstadter, R., Hughes, E. B., Hunter, S. D., Hughlock, B. W., 
et al. 1993, ApJS, 86, 629. 

Thompson, D. J., Bertsch, D. L., Dingus, B. L., Esposito, J. A., 
Etienne, A., Fichtel, C. E., Friedlander, D. P., Hartman, R. C., et al. 
1995 ApJS, 101, 259. 

Totani, T. \& Kitayama, T. 2000, ApJ, 545, 572. 

Tucker, W. H., Tananbaum, H., Remillard, R. A. 1995, ApJ, 444, 532. 

Turner, T. J., George, I. M., Nandra, K., Mushotzky, R. F. 1997, ApJS, 113, 
23. 

Ueda, Y., Ishisaki, Y., Takahashi, T., Makishima, K., Ohashi, T. 2001, ApJS,
133, 1. 

Urry, C. M., Padovani, P. 1995, PASP, 107, 803. 

Voges, W., Aschenbach, B., Boller, Th., Br\"auninger, H., Briel, U., et al. 
2000, VizieR On-line Data Catalog: IX/29. Originally published in:
Max-Planck-Institut fur extraterrestrische Physik, Garching (2000). 

Voges, W., Aschenbach, B., Boller, Th., Br\"auninger, H., Briel, U., et al. 
1999, A\&A, 349, 389. 

Watanabe, K. \& Hartmann, D. H. 2001, Proc. GAMMA 2001, Baltimore;
astro-ph/0105180.  

Weekes, T. C.,  Bradbury, S. M., Bond, I. H., Breslin, A. C., Buckley, J. H., 
et al., 2000, The Fifth Compton Symposium, Proceedings of the fifth Compton 
Symposium, held in Portsmouth, NH, USA, September 1999. Melville, NY: 
American Institute of Physics (AIP), 2000. Edited by M. L. McConnell and
J. M. Ryan AIP Conference Proceedings, Vol. 510., p.637.  

Weferling, B., \& Schlickeiser, R. 1999, A\&A, 344, 744. 

Welty, A. D. \& Ramsey, W. 1995, ApJ, 109, 2187. 

Worrall, M. \& Birkinshaw, D. M. 1994, ApJ, 427, 134. 

\end{document}